\global\def\draftcontrol{0}

   \def\versionno{border}

\catcode`\@=11

\expandafter\ifx\csname draftcontrol\endcsname\relax\global\def\draftcontrol{0}
\fi

{\count255=\time\divide\count255 by 60
\xdef\hourmin{\number\count255}
\multiply\count255 by-60\advance\count255 by\time
\xdef\hourmin{\hourmin:\ifnum\count255<10 0\fi\the\count255}}
\def\draftdate{\number\month/\number\day/\number\year\ \ \ \hourmin }

\newcommand\makepapertitle{\par
  \begingroup
    \renewcommand\thefootnote{\@fnsymbol\c@footnote}%
    \def\@makefnmark{\rlap{\@textsuperscript{\normalfont\@thefnmark}}}%
    \long\def\@makefntext##1{\parindent 1em\noindent
            \hb@xt@1.8em{%
                \hss\@textsuperscript{\normalfont\@thefnmark}}##1}%
     \newpage
     \global\@topnum\z@   
     \@makepapertitle
     \thispagestyle{empty}\@thanks
  \endgroup
  \setcounter{footnote}{0}%
  \global\let\thanks\relax
  \global\let\makepapertitle\relax
  \global\let\@makepapertitle\relax
  \global\let\@thanks\@empty
  \global\let\@author\@empty
  \global\let\@date\@empty
  \global\let\@title\@empty
  \global\let\title\relax
  \global\let\author\relax
  \global\let\date\relax
  \global\let\and\relax
  \def\version{\let\version\@version\@gobble}
}
\def\@makepapertitle{%
  \newpage
   \ifnum\draftcontrol=1 {}
   \version\versionno
   \vskip 3em%
   \else
   \hfill\hbox to 3cm {\parbox{4cm}{\@pubnum}\hss}%
   \vskip 3em%
   \fi
   \begin{center}%
   \let \footnote \thanks
     {\LARGE {\@title}}%
     \vskip 1.5em%
     {\normalsize
       \lineskip .5em%
       \begin{tabular}[t]{c}%
         \@author
       \end{tabular}\par}%
     \vskip 1.5em%
     {\@bstract}%
     \end{center}%
     \vskip 1.5em
     \@date%
   \par
}

\gdef\@pubnum{}
\def\pubnum#1{%
  \gdef\@pubnum{#1}}

\gdef\@bstract{}
\def\Abstract#1{%
  \gdef\@bstract{%
   \parbox{\textwidth-0pc}{%
   \centerline{\bf Abstract}\penalty1000%
\kern.2cm%
\noindent
\renewcommand\baselinestretch{1.0}%
{#1}}}
}

\def\ps@paper{\let\@mkboth\@gobbletwo%
     \ifnum\draftcontrol=1
    \def\@oddfoot{\hbox to \textwidth{\tiny \versionno \hfil\tiny\draftdate}%
    \hskip -\textwidth \hbox to \textwidth{\hfil\rm\thepage\hfil}}%
     \else\def\@oddfoot{\hbox to \textwidth{\hfil\rm\thepage\hfil}}
     \fi
     \let\@evenfoot\@oddfoot
}

\def\body{\clearpage
          \pagestyle{paper}
    }

\def\@version#1{\ifnum\draftcontrol=1
\typeout{}\typeout{#1}\typeout{}
\vskip3mm\centerline{\hbox{\fbox{\normalsize{\tt DRAFT -- #1 -- }
                   {\draftdate}}}}\vskip3mm
\fi}
\let\version\@version
\long\def\eqlabel#1{\ifnum\draftcontrol=1
                    \tag@false  
                    \tag*{(\theequation) \hbox to -0.2cm{\hspace{0cm}\small{#1}\hss}}
                    \refstepcounter{equation}
                    \edef\@currentlabel{\theequation}
                    \ltx@label{#1}          
                    \else
                    \label{#1}
                    \fi
                    }
\let\st@bibitem\@bibitem
\let\st@lbibitem\@lbibitem
\ifnum\draftcontrol=1
  \def\@bibitem#1{%
    \st@bibitem{#1}\a@@label{#1}\ignorespaces}
  \def\@lbibitem[#1]#2{%
    \st@lbibitem[#1]{#2}\a@@label{#2}\ignorespaces}
  \def\a@@label#1{%
    \gdef\a@lab{\smash{\normalfont\small#1}}
    \ifvmode
      \if@inlabel
        \global\setbox\@labels\hbox{%
          \llap{\a@lab\let\a@lab\relax
                \kern\@totalleftmargin\kern\marginparsep}%
          \box\@labels}%
      \fi
    \fi}
\fi

\documentclass[12pt,letterpaper]{article}

\usepackage{amsmath,amssymb,array,calc,epsfig,rotating,bm}
\usepackage[sort]{cite}
\usepackage{graphicx}
\usepackage{psfrag,verbatim}
\usepackage{xcolor}

\ifnum\draftcontrol=1
\fi

\renewcommand\baselinestretch{1.25}
\renewcommand\section{\@startsection {section}{1}{\z@}%
                                   {-3.5ex \@plus -1ex \@minus -.2ex}%
                                   {2.3ex \@plus.2ex}%
                                   {\normalfont\large\bfseries}}
\renewcommand\subsection{\@startsection{subsection}{2}{\z@}%
                                   {-3.25ex\@plus -1ex \@minus -.2ex}%
                                   {1.5ex \@plus .2ex}%
                                   {\normalfont\normalsize\bfseries}}
\renewcommand\subsubsection{\@startsection{subsubsection}{3}{\z@}%
                                   {-3.25ex\@plus -1ex \@minus -.2ex}%
                                   {1.5ex \@plus .2ex}%
                                   {\normalfont\normalsize\it}}
\renewcommand\paragraph{\@startsection{paragraph}{4}{\z@}%
                                   {-3.25ex\@plus -1ex \@minus -.2ex}%
                                   {1.5ex \@plus .2ex}%
                                   {\normalfont\normalsize\bf}}

\def\revise#1       {\raisebox{-0em}{\rule{3pt}{1em}}%
                     \marginpar{\raisebox{.5em}{\vrule width3pt\
                     \vrule width0pt height 0pt depth0.5em
                     \hbox to 0cm{\hspace{0cm}{%
                     \parbox[t]{4em}{\raggedright\footnotesize{#1}}}\hss}}}}

\newcommand{\ie}{{\it i.e.,}\ }
\newcommand{\eg}{{\it e.g.,}\ }

\def\cale         {{\cal E}}

\def\call         {{\cal L}}

\def\caln         {{\cal N}}
\def\calo         {{\cal O}}
\def\calp         {{\cal P}}

\def\zet          {{\mathbb Z}}

\def\del          {\partial}

\def\sqr#1#2{{\vcenter{\vbox{\hrule height.#2pt
 \hbox{\vrule width.#2pt height#1pt \kern#1pt
 \vrule width.#2pt}\hrule height.#2pt}}}}

\def\aa1{\phi}
\def\cc1{\psi}

\catcode`\@=12

\begin{document}


\title{\bf SUGRA/Strings like to be bald}

\date{July 18, 2020}

\author{
Alex Buchel \\
\it Department of Applied Mathematics\\
\it Department of Physics and Astronomy\\ 
\it University of Western Ontario\\
\it London, Ontario N6A 5B7, Canada\\
\it Perimeter Institute for Theoretical Physics\\
\it Waterloo, Ontario N2J 2W9, Canada\\[0.4cm]
}

\Abstract{We explore the embedding of the phenomenological holographic models
describing thermal relativistic ordered conformal phases in ${\mathbb
R}^{2,1}$ in SUGRA/String theory.  The dual black branes in a Poincare
patch of asymptotically $AdS_4$ have ``hair'' --- a condensate of the
order parameter for the broken symmetry. In a gravitational dual the
order parameter for a spontaneous symmetry breaking is represented by
a bulk scalar field with a nontrivial potential. To construct the
ordered conformal phases perturbatively we introduce a
phenomenological deformation parameter in the scalar potential. We
find that while the ordered phases exist for different values of the
deformation parameter, they disappear before the deformation is
removed, in one case once the potential is precisely as in the
top-down holography.  It appears that the holographic models with the
conformal ordered phases are in the String theory swampland.
}

\makepapertitle

\body

\version\versionno

Thermal relativistic ordered conformal phases are states of a $d$-dimensional conformal field theory,
$CFT_d$, in ${\mathbb R}^{d-1,1}$ with spontaneously broken global symmetry.
They are characterized by the energy density $\cale$, the pressure $P$, the entropy density $s$, and an expectation
value of an operator (one or more) of a conformal dimension $\Delta$, $\calo_\Delta$, representing the
condensate of the order parameter: 
\begin{equation}
\cale\propto T^d\,,\qquad P=\frac{1}{d-1}\ \cale\,,\qquad s=\frac{d}{d-1}\ \frac{\cale}{T}\,,\qquad
\calo_\Delta\propto T^\Delta\,.
\eqlabel{ophase}
\end{equation}
Conformal ordered phases from the QFT perspective were recently discussed in \cite{Chai:2020zgq},
and their particular holographic realization was proposed in
\cite{Buchel:2020thm}\footnote{See \cite{Buchel:2009ge} for related nonconformal models.}.
The holographic models of \cite{Buchel:2020thm} are phenomenological, and in this paper we explore
the question whether thermal conformal ordered phases can arise in top-down SUGRA/String theory holographic models. 

In a gravitational dual, conformal ordered phases are represented by  black branes in a
Poincare patch of asymptotically $AdS_{d+1}$ bulk with a nonlinear scalar hair. This nonlinearity makes the
construction rather subtle. We present a procedure\footnote{The method is very general
and can be applied to any holographic model.} where a scalar potential of the top-down
holographic model is deformed with a parameter $b$, akin to the $b$-parameter in  phenomenological
models of \cite{Buchel:2020thm}, such that at $b=1$ we restore
the exact SUGRA scalar gravitational potential. In the limit $b\to +\infty$ the 'hairy' black branes
can be constructed perturbatively as small deformations of the AdS-Schwarzschild black brane.
As in  \cite{Buchel:2020thm}, we find that the holographic conformal ordered phases exist
for $b\in (b_{crit},+\infty)$. In the limit $b\to b_{crit}+0$ the order parameter
diverges \cite{Buchel:2020thm}. In all cases discussed, we find that $b_{crit}\ge 1$. Thus, it appears that 
holographic models realizing ordered conformal phases are in the String theory swampland
\cite{Vafa:2005ui}.

We consider two specific holographic models in $\caln=8$ gauged supergravity with the
bulk scalars dual to bosonic/fermionic bilinears in the M2 brane theory:
\begin{itemize}
\item {\bf Model A:}  is a consistent truncation of four-dimensional $\caln=8$ gauged supergravity
to $U(1)^4$  invariant sector  \cite{Chong:2004ce,Donos:2011ut}. Here we have for the scalar
potential 
\begin{equation}
\calp_A=-2g^2\ \biggl [1+2\cosh^2\phi\biggr] \,.
\eqlabel{pa}
\end{equation}
Notice that $\calp_A$ is unbounded from below.
\item {\bf Model B:}  is a consistent truncation of four-dimensional $\caln=8$ gauged supergravity
to $SO(3)\times SO(3)$ invariant sector  \cite{Bobev:2011rv}. Here we have for the scalar
potential\footnote{This fixes the typo in eq.(2.11) in \cite{Bobev:2011rv}. I would like to thank
Nikolay Bobev for valuable correspondence.}
\begin{equation}
\calp_B=-\frac {g^2}{2}\ \biggl [3+10\cosh^2\phi -\cosh^4\phi\biggr] \,.
\eqlabel{pb}
\end{equation}
Unlike $\calp_A$, the scalar potential $\calp_B$ is bounded from below. 
\end{itemize}
We take the Lagrangian of the scalars coupled to gravity to be:
\begin{equation}
\call=\frac 12 R-(\del\phi)^2-\calp\,.
\eqlabel{lag}
\end{equation}
In both cases $g$ is a SUGRA coupling constant; we set $2g^2=1$, which sets the radius of the asymptotic
$AdS_4$ geometry to one. Note that 
\begin{equation}
\begin{split}
\calp_A=&-3 -2 \phi^2-\frac 23\phi^4-\calo(\phi^6)\,,\\
\calp_B=&-3 -2 \phi^2-\frac{5}{12}\phi^4+\calo(\phi^6)\,,
\end{split}
\eqlabel{series}
\end{equation}
thus, the bulk scalar $\phi$ is dual to an operator of $\Delta=1$ or $\Delta=2$ ---
we will discuss both allowed quantizations. 
Both $\calp_A$ and $\calp_B$ are invariant under $\phi\leftrightarrow -\phi$; so a nonzero expectation
value of $\calo_\Delta$ signals a spontaneous breaking\footnote{We will always keep nonzero only one of
the two normalizable modes of $\phi$ near the boundary.} of this $\zet_2$ symmetry.

To study the thermal phases of \eqref{lag}, we take the following background ansatz
\begin{equation}
ds_4^2=-c_1^2\ dt^2 +c_2^2\ \left[dx_1^2+dx_2^2\right]+ c_3^2\ dr^2\,,
\eqlabel{bbs}
\end{equation}
where all the metric warp factors $c_i$ as well as the bulk scalar $\phi$ are functions of the
radial coordinate $r$,
\begin{equation}
r\ \in\ [r_0,+\infty)\,,
\eqlabel{rrange}
\end{equation}
where $r_0$ is a location of a regular Schwarzschild horizon, and $r\to +\infty$
is the asymptotic $AdS_4$ boundary. Introducing a new radial coordinate
\begin{equation}
x\equiv \frac{r_0}{r}\,,\qquad x\ \in (0,1]\,,
\eqlabel{defx}
\end{equation}
and denoting
\begin{equation}
\begin{split}
&c_1=r \left(1-\frac{r_0^3}{r^3}\right)^{1/2}\ a_1\,,\qquad c_2=r\,,\qquad
c_3=\frac 1r\ \left(1-\frac{r_0^3}{r^3}\right)^{-1/2}\ a_3\,,
\end{split}
\eqlabel{defcs}
\end{equation}
we obtain the following system of ODEs (in a radial coordinate $x$, $'=\frac{d}{dx}$,
$\del=\frac{\delta}{\delta\phi}$):
\begin{equation}
\begin{split}
&0=a_1'+\frac{3a_1}{2x (x^3-1)}+\frac12 x a_1 \left(\phi'\right)^2
+\frac{a_3^2 a_1}{2x (x^3-1)}\ \calp\,,
\end{split}
\eqlabel{eq1}
\end{equation}
\begin{equation}
\begin{split}
&0=a_3'+\frac12 x a_3 \left(\phi'\right)^2
-\frac{3a_3}{2x (x^3-1)}-\frac{a_3^3}{2x (x^3-1)}\ \calp\,,
\end{split}
\eqlabel{eq2}
\end{equation}
\begin{equation}
\begin{split}
&0=\phi''+\left(\frac{1}{x}+\frac{a_3^2\ \calp}{x (1-x^3)}\right)
\phi'+\frac{a_3^2}{2(x^3-1) x^2}\ \del \calp\,.
\end{split}
\eqlabel{eq3}
\end{equation}

There is always a trivial solution to \eqref{eq1}-\eqref{eq3}, \ie
\begin{equation}
a_1\equiv 1\,,\qquad a_3\equiv 1\,,\qquad \phi\equiv 0\,,
\eqlabel{sym}
\end{equation}
corresponding to a disordered, $\zet_2$ symmetric phase,
\begin{equation}
\cale=\frac{c}{96}\ (\pi T)^2\,,\qquad \calo_\Delta=0 \,,
\eqlabel{sym1}
\end{equation}
where $c$ is a central charge of the boundary $CFT_3$.

We are after the nontrivial solution to \eqref{eq1}-\eqref{eq3},
with $\phi(x)\ne 0$ --- in this phase the parity symmetry is spontaneously broken.
Rather than to study ordered phases with a scalar potential $\calp$, we introduce
a deformed scalar potential $\calp^b$, for which the existence of the broken phases,
at least in the limit $b\to +\infty$, is physically well motivated:
\begin{equation}
\calp^b[\phi]\equiv -3 -2\phi^2 +b\biggl(\calp[\phi]+3+2\phi^2\biggr)\,,\qquad \calp^{b=1}[\phi]\equiv
\calp[\phi]\,.
\eqlabel{defpb}
\end{equation}
Indeed, as $b\to \infty$ there is a systematic  power series solution of
\eqref{eq1}-\eqref{eq3}:
\begin{equation}
\begin{split}
&\phi=\frac{1}{b^{1/2}}\ \phi_1+\frac{1}{b^{3/2}}\ \phi_3+\calo(b^{-3/2})\,,\qquad
a_{1}=1+\frac 1b\ a_{1,1}+\frac{1}{b^2}\ a_{1,2}+\calo(b^{-2})\,,\\
&a_{3}=1+\frac 1b\ a_{3,1}+\frac{1}{b^2}\ a_{3,2}+\calo(b^{-2})\,.
\end{split}
\end{equation}
Notice that to leading order, \eg
\begin{equation}
\calp_A^b=-3 -2\phi^2 -b\ \frac 23\ \phi^4+b\ \calo(\phi^6)=-3
+\underbrace{m_{eff}^2\  \phi^2}_{\calo(b^{-1})}+\calo(b^{-3})  \,,
\end{equation}
with
\begin{equation}
m_{eff}^2=\underbrace{\Delta(\Delta-3)}_{-2}-b \frac 23 \phi^2 =
\underbrace{\Delta(\Delta-3)-\frac 23 \phi_1^2}_{\calo(b^0)}
+\calo(b^{-1})\,.
\eqlabel{defmeff}
\end{equation}
That is, to leading order in the limit $b\to \infty$, the effecting
mass of the bulk scalar $\phi$ is shifted due to a nonlinear negative quartic term in
$\calp^b$. Potentially, when evaluated at the AdS-Schwarzschild
horizon\footnote{To leading order in $b$ the bulk geometry is not modified.}, it can 
dip below the Breitenlohner-Freedman bound, triggering the instability and leading to 'hair'.

To check whether or not the hair indeed arises, one needs to perform  an explicit computation.
We focus\footnote{The other cases can be considered verbatim.}
on the model with a scalar potential $\calp_A^b$. To leading order as $b\to \infty$ we find
\begin{equation}
0=\phi_1''+\frac{x^3+2}{x(x^3-1)}\ \phi_1' -\frac{2\phi_1(2\phi_1^2+3)}{3(x^3-1)x^2}\,.
\eqlabel{eomphi1}
\end{equation}
If we identify the bulk scalar as dual to the operator of $\Delta=1$, we must impose the
boundary asymptotics on
$\phi_1$ as
\begin{equation}
\begin{split}
&{\rm UV}:\qquad \phi_1= f_{1,1}\ x+\calo(x^3)\,,\qquad x\to 0\,,\\
&{\rm IR}:\qquad \phi_1=f_{1,0}^h+\calo((1-x)^1)\,,\qquad (1-x)\to 0\,.
\end{split}
\eqlabel{asphi1}
\end{equation}
A nontrivial solution of \eqref{eomphi1} with \eqref{asphi1} can be used to construct
perturbative hair, a $\zet_2$ broken phase in the limit $b\to +\infty$. In all models we considered,
provided the quartic term in the scalar potential $\calp$ is negative, such a nontrivial solution
indeed exists\footnote{In fact, there is a whole 'tower' --- a discrete spectrum --- of solutions.}.
We find
\begin{equation}
f_{1,1}=\pm 0.6369\,,\qquad f_{1,0}^h=\pm 1.10602\,.
\eqlabel{resphi1}
\end{equation}
Once the solution for $\phi_1$ is constructed, it will source the linear equations for $a_{1,1}$
and $a_{3,1}$. The conformal black brane with a perturbative hair thus constructed can be numerically
extended to finite values of $b$, realizing the construction of the holographic dual for a
conformal order in the $b$-deformed model.
Details of the construction follow \cite{Buchel:2020thm} and will
be omitted here. We parameterize the ordered conformal equation of state as
\begin{equation}
\cale= \frac{c}{96}\ (\pi T)^2\ \times\ \kappa_A^{\Delta}(b)\,,\qquad \frac{\calo_{\Delta,A}}{T^\Delta}\equiv
\hat\calo_{\Delta,A}(b)\,.
\eqlabel{eos}
\end{equation}

\begin{figure}[t]
\begin{center}
\psfrag{b}[cc][][1.0][0]{$b$}
\psfrag{k}[cc][][1.0][0]{$\kappa_A^{\Delta=1}$}
\psfrag{o}[cc][][1.0][0]{$|\hat{\calo}_{1,A}|^{-2}$}
\includegraphics[width=3in]{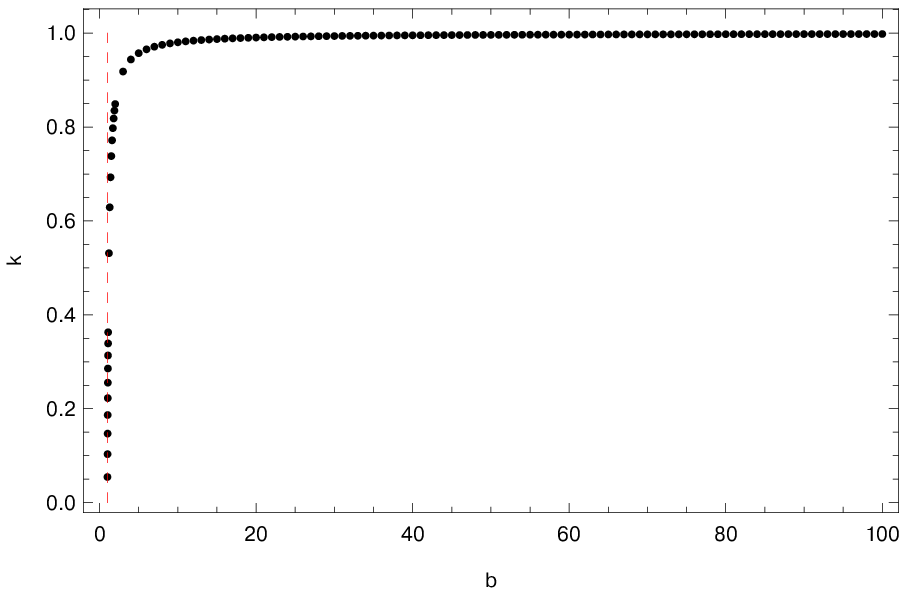}\,
\includegraphics[width=3in]{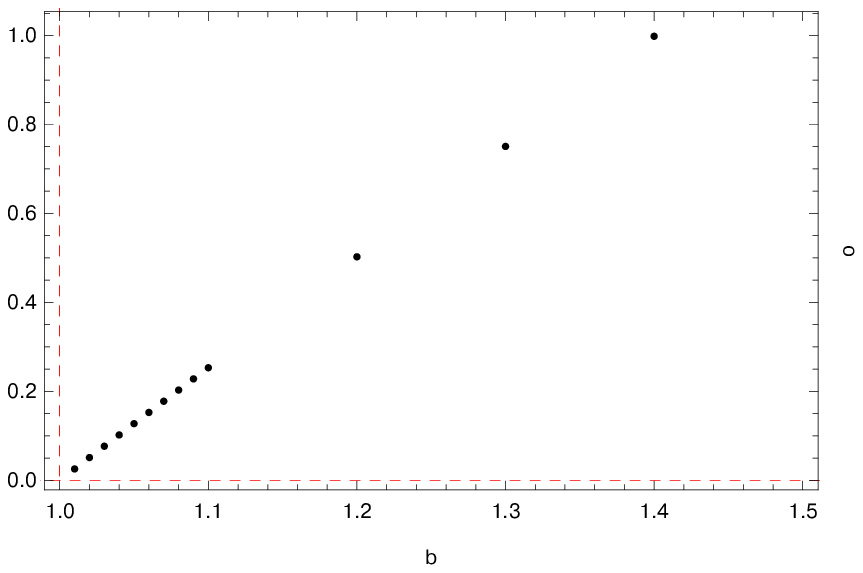}
\end{center}
  \caption{Holographic conformal order in {\bf Model A} with a $b$-deformed scalar potential
  $\calp_A^b$ and the bulk scalar quantization $\Delta=1$.
  Parameter $\kappa_A^{\Delta=1}$ defines the equation of state, see \eqref{eos} (left panel).
  Right panel: the order parameter $\hat{\calo}_{1,A}$ for the spontaneous $\zet_2$ symmetry
  breaking, see \eqref{eos}. Vertical dashed red line indicates extrapolated value of
  $b_{crit}^{A,\Delta=1}=1$ at which the order parameter diverges. The ordered phase does not exist for
  $b\le b_{crit}^{A,\Delta=1}$.
} \label{fig1}
\end{figure}

In fig.~\ref{fig1} we present results for $\kappa_A^{\Delta=1}$ as a function of $b$ for a holographic model
with a scalar potential $\calp_A^b$. The ordered phase exists only for $b>1$, and as $b\to 1$,
the order parameter for the spontaneous symmetry breaking diverges. In practice we constructed
conformal order with $b\ge 1.01$; extrapolation of the data for the
order parameter predicts its divergence at $b\approx 0.999997$. We take it as a strong indication that
the critical value of $b$, below which the ordered phase does not exist, is
\begin{equation}
b_{crit}^{A,\Delta=1}=1\,.
\eqlabel{bcrita1}
\end{equation}
Remarkable,  the critical value of $b$ is just what is needed for the model with $\calp_A^b$
to become a top-down holography. We conclude that Model A with a bulk scalar field quantization
$\Delta=1$ does not admit a thermal phase
with spontaneously broken $\zet_2$ symmetry.

\begin{figure}[t]
\begin{center}
\psfrag{b}[cc][][1.0][0]{$b$}
\psfrag{k}[cc][][1.0][0]{$\kappa_A^{\Delta=2}$}
\psfrag{o}[cc][][1.0][0]{$|\hat{\calo}_{2,A}|^{-1}$}
\includegraphics[width=3in]{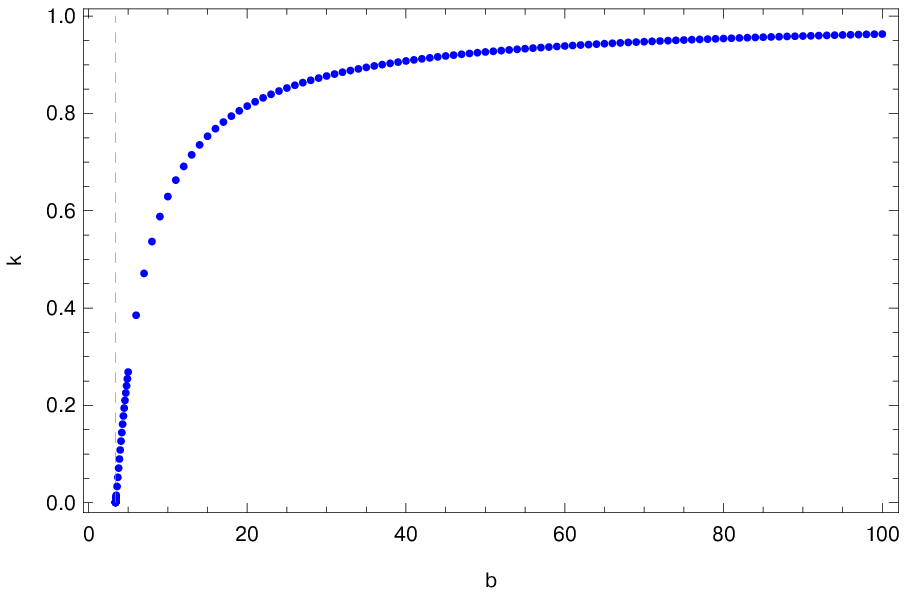}\,
\includegraphics[width=3in]{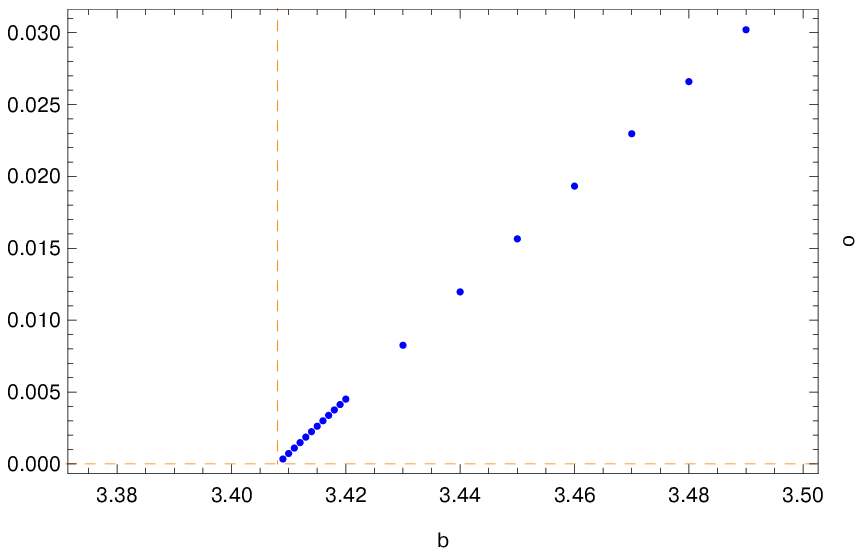}
\end{center}
  \caption{Holographic conformal order in {\bf Model A} with a $b$-deformed scalar potential
  $\calp_A^b$ and the bulk scalar quantization $\Delta=2$.
  Parameter $\kappa_A^{\Delta=2}$ defines the equation of state  (left panel).
  Right panel: the order parameter $\hat{\calo}_{2,A}$ for the spontaneous $\zet_2$ symmetry
  breaking. Vertical dashed orange line indicates extrapolated value of
  $b_{crit}^{A,\Delta=2}=3.4081$ at which the order parameter diverges. The ordered phase does not exist for
  $b\le b_{crit}^{A,\Delta=2}$.
} \label{fig2}
\end{figure}

As shown in fig.~\ref{fig2}, Model A with a  bulk scalar field quantization
$\Delta=2$ does not admit a thermal phase
with spontaneously broken $\zet_2$ symmetry either, since  $b_{crit}^{A,\Delta=2}>1$.

\begin{figure}[t]
\begin{center}
\psfrag{b}[cc][][1.0][0]{$b$}
\psfrag{k}[cc][][1.0][0]{$\kappa_B^{\Delta=1}$}
\psfrag{o}[cc][][1.0][0]{$|\hat{\calo}_{1,B}|^{-4}$}
\includegraphics[width=3in]{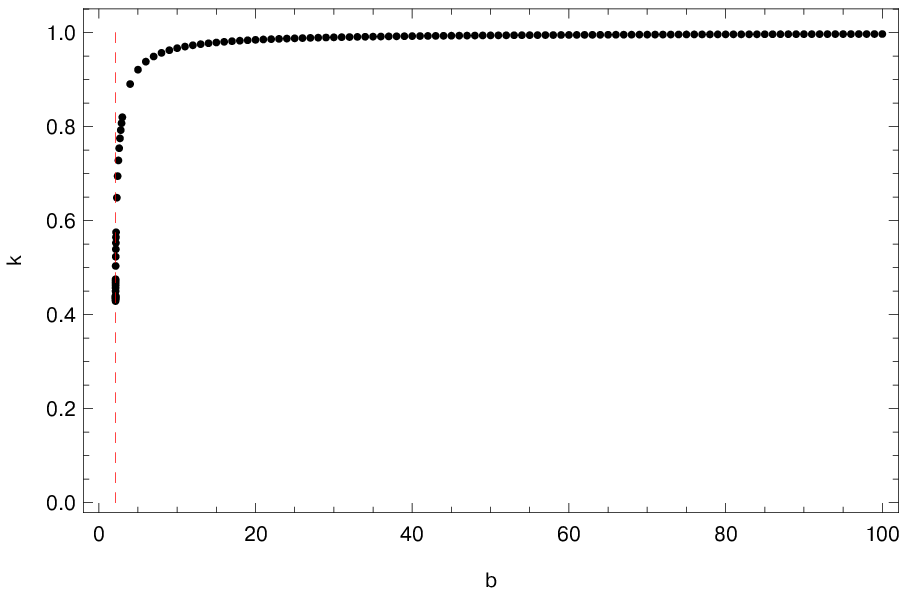}\,
\includegraphics[width=3in]{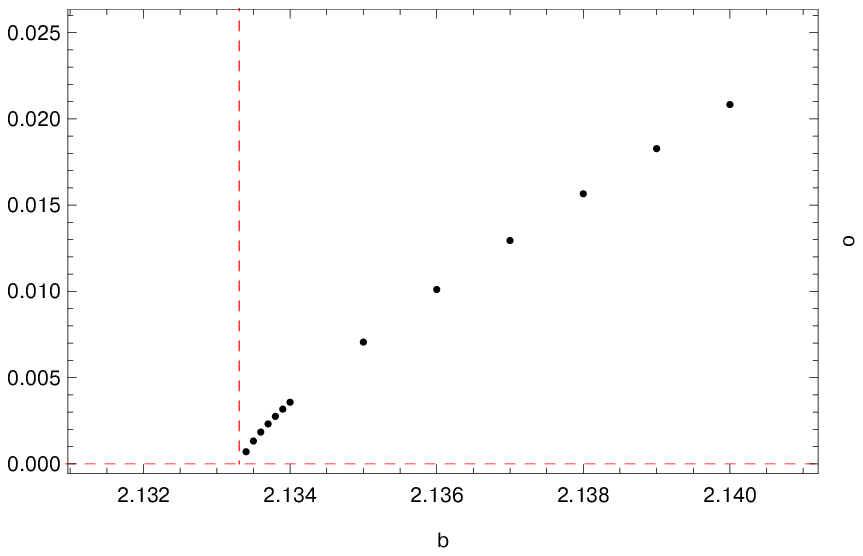}
\end{center}
  \caption{Holographic conformal order in {\bf Model B} with a $b$-deformed scalar potential
  $\calp_B^b$ and the bulk scalar quantization $\Delta=1$.
  Parameter $\kappa_B^{\Delta=1}$ defines the equation of state  (left panel).
  Right panel: the order parameter $\hat{\calo}_{1,B}$ for the spontaneous $\zet_2$ symmetry
  breaking. Vertical dashed red line indicates extrapolated value of
  $b_{crit}^{B,\Delta=1}=2.13331$ at which the order parameter diverges. The ordered phase does not exist for
  $b\le b_{crit}^{B,\Delta=1}$.
} \label{fig3}
\end{figure}

\begin{figure}[t]
\begin{center}
\psfrag{b}[cc][][1.0][0]{$b$}
\psfrag{k}[cc][][1.0][0]{$\kappa_B^{\Delta=2}$}
\psfrag{o}[cc][][1.0][0]{$|\hat{\calo}_{2,B}|^{-2}$}
\includegraphics[width=3in]{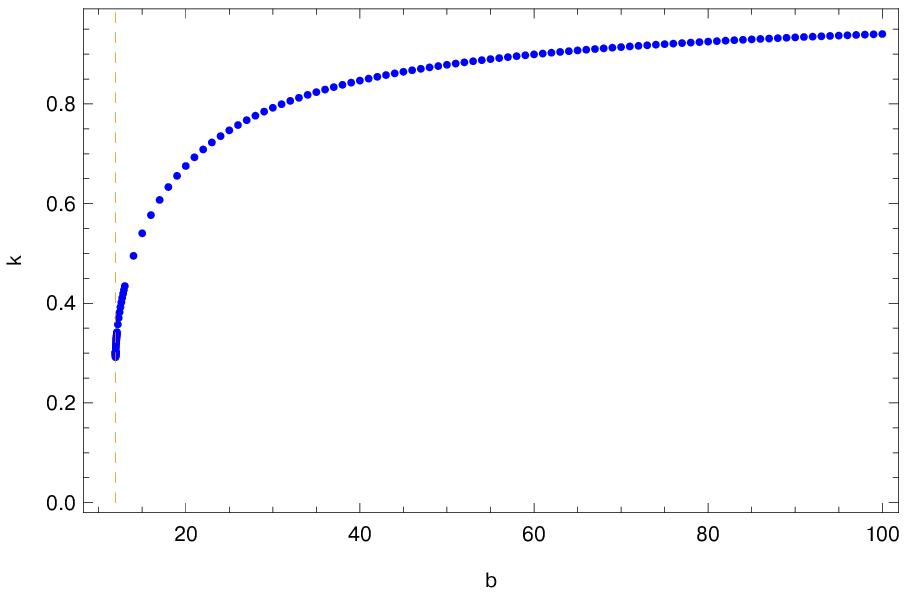}\,
\includegraphics[width=3in]{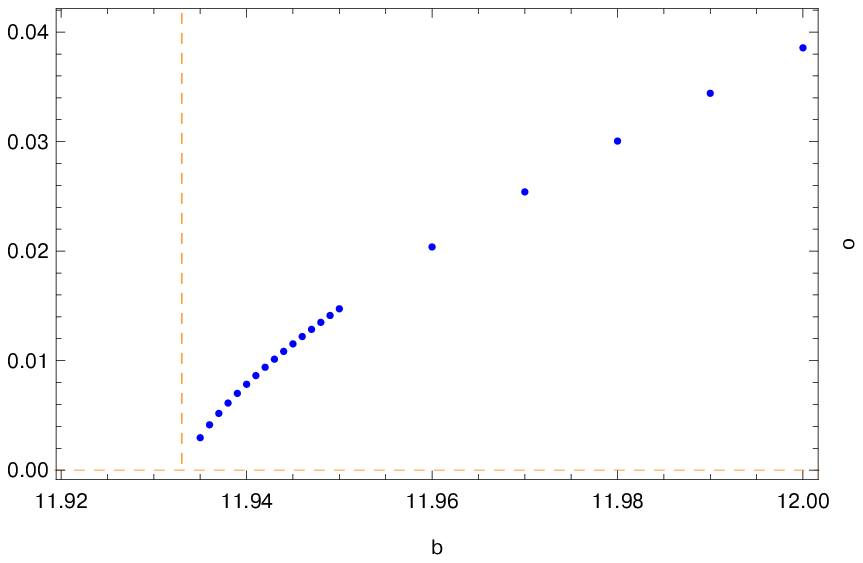}
\end{center}
  \caption{Holographic conformal order in {\bf Model B} with a $b$-deformed scalar potential
  $\calp_B^b$ and the bulk scalar quantization $\Delta=2$.
  Parameter $\kappa_B^{\Delta=2}$ defines the equation of state  (left panel).
  Right panel: the order parameter $\hat{\calo}_{2,B}$ for the spontaneous $\zet_2$ symmetry
  breaking. Vertical dashed orange line indicates extrapolated value of
  $b_{crit}^{B,\Delta=2}=11.933$ at which the order parameter diverges. The ordered phase does not exist for
  $b\le b_{crit}^{B,\Delta=2}$.
} \label{fig4}
\end{figure}

In figs.~\ref{fig3}-\ref{fig4} we report the corresponding results for Model B,
for $\Delta=1$ and $\Delta=2$ quantizations of the bulk scalar field. Once again, since 
$b_{crit}^{B,\Delta=1}>1$ and $b_{crit}^{B,\Delta=2}>1$, there are no conformal ordered phases in Model B.

\section*{Acknowledgments}
Research at Perimeter
Institute is supported by the Government of Canada through Industry
Canada and by the Province of Ontario through the Ministry of
Research \& Innovation. This work was further supported by
NSERC through the Discovery Grants program.\bibliographystyle{JHEP}

\bibliography{border}

\providecommand{\href}[2]{#2}\begingroup\raggedright\begin{thebibliography}{1}

\bibitem{Chai:2020zgq}
N.~Chai, S.~Chaudhuri, C.~Choi, Z.~Komargodski, E.~Rabinovici and M.~Smolkin,
  \emph{{Thermal Order in Conformal Theories}},
  \href{https://arxiv.org/abs/2005.03676}{{\tt 2005.03676}}.

\bibitem{Buchel:2020thm}
A.~Buchel, \emph{{Thermal order in holographic CFTs and no-hair theorem
  violation in black branes}},  \href{https://arxiv.org/abs/2005.07833}{{\tt
  2005.07833}}.

\bibitem{Buchel:2009ge}
A.~Buchel and C.~Pagnutti, \emph{{Exotic Hairy Black Holes}},
  \href{http://dx.doi.org/10.1016/j.nuclphysb.2009.08.017}{\emph{Nucl. Phys. B}
  {\bf 824} (2010) 85--94}, [\href{https://arxiv.org/abs/0904.1716}{{\tt
  0904.1716}}].

\bibitem{Vafa:2005ui}
C.~Vafa, \emph{{The String landscape and the swampland}},
  \href{https://arxiv.org/abs/hep-th/0509212}{{\tt hep-th/0509212}}.

\bibitem{Chong:2004ce}
Z.-W. Chong, H.~Lu and C.~Pope, \emph{{BPS geometries and AdS bubbles}},
  \href{http://dx.doi.org/10.1016/j.physletb.2005.03.050}{\emph{Phys. Lett. B}
  {\bf 614} (2005) 96--103}, [\href{https://arxiv.org/abs/hep-th/0412221}{{\tt
  hep-th/0412221}}].

\bibitem{Donos:2011ut}
A.~Donos and J.~P. Gauntlett, \emph{{Superfluid black branes in $AdS_4 \times
  S^7$}}, \href{http://dx.doi.org/10.1007/JHEP06(2011)053}{\emph{JHEP} {\bf 06}
  (2011) 053}, [\href{https://arxiv.org/abs/1104.4478}{{\tt 1104.4478}}].

\bibitem{Bobev:2011rv}
N.~Bobev, A.~Kundu, K.~Pilch and N.~P. Warner, \emph{{Minimal Holographic
  Superconductors from Maximal Supergravity}},
  \href{http://dx.doi.org/10.1007/JHEP03(2012)064}{\emph{JHEP} {\bf 03} (2012)
  064}, [\href{https://arxiv.org/abs/1110.3454}{{\tt 1110.3454}}].

\end{thebibliography}\endgroup

\end{document}